\documentclass{emulateapj}
\newcommand{\grad}{\mbox{$\rlap.{^{\mathrm{o}}}$}}      
\slugcomment{Accepted by The Astrophysical Journal}
\shortauthors{USCANGA, CANT\'O, \& RAGA}

\begin{document}
\title{Position--Velocity Diagrams for the Maser Emission coming from a Keplerian Ring}

\author{Lucero Uscanga,}
\affil{Centro de Radioastronom\'{\i}a y Astrof\'{\i}sica, Universidad Nacional Aut\'onoma de M\'exico}
\affil{Apartado Postal 3-72, 58089 Morelia, Michoac\'an, Mexico}
\email{l.uscanga@astrosmo.unam.mx}

\author{Jorge Cant\'o,}
\affil{Instituto de Astronom\'{\i}a, Universidad Nacional Aut\'onoma
de M\'exico}
\affil{Apartado Postal 70-264, 04510 M\'exico, DF, Mexico}

\and

\author{Alejandro C. Raga}
\affil{Instituto de Ciencias Nucleares, Universidad Nacional Aut\'onoma de M\'exico}
\affil{Apartado Postal 70-543, 04510 M\'exico, DF, Mexico}
\email{raga@nucleares.unam.mx}

\begin{abstract}
We have studied the maser emission from a thin, planar, gaseous ring in Keplerian rotation around a central mass observed edge-on. The absorption coefficient within the ring is assumed to follow a power law dependence with the distance from the central mass as, $\kappa=\kappa_0r^{-q}$. We have calculated position-velocity diagrams for the most intense maser features, for different values of the exponent $q$. We have found that, depending on the value of $q$, these diagrams can be qualitatively different. The most intense maser emission at a given velocity can either come mainly from regions close to the inner or outer edges of the amplifying ring or from the line perpendicular to the line of sight and passing through the central mass (as is commonly assumed). Particularly, when $q>1$ the position-velocity diagram is qualitatively similar to the one observed for the water maser emission in the nucleus of the galaxy NGC 4258. In the context of this simple model, we conclude that in this object the absorption coefficient depends on the radius of the amplifying ring as a decreasing function, in order to have significant emission coming from the inner edge of the ring.
\end{abstract}


\section{Introduction}
The a priori probability of seeing a thin disk nearly edge-on is very small. It is given by $p\simeq 0.125\,(h/R)^2$, where $h$ is the thickness of the disk and $R$ is its radius. Typically $h/R\simeq 0.01$ and thus $p\simeq 1.25\times 10^{-5}$. Surprisingly however, the maser emission observed in several cosmic sources has been successfully modeled as coming from a ring or truncated disk in Keplerian rotation (around a massive object) seen edge-on. For instance: circumstellar disks in star-forming regions as in S255 \citep{Ces90} and MWC 349 \citep{Pon94}, and also circumnuclear disks around black holes of galactic nuclei as in NGC 4258 \citep{Wat94,Miy95}. In general, the maser emission from a Keplerian disk observed edge-on produces a triple-peaked spectrum \citep{Elm79}; but \citet{Pon94} showed that, there is a transition from triple- to double-peaked spectra as the width of the amplifying ring decreases.

NGC 4258 is a Seyfert 2/LINER located at a distance of $7.2\pm 0.3$~Mpc \citep{Her99}. The water maser emission (22 GHz) toward this galaxy was first detected by \citet{Cla84}. Shortly afterwards it was shown that the water masers are confined in a very small region ($\sim$1.3~pc) at the center of NGC 4258 \citep{Cla86}. Subsequently, \citet{Nak93} discovered water maser emission with velocity offsets $\pm$1000~km~s$^{-1}$ from the already known emission at the galactic systemic velocity of $\simeq$472~km~s$^{-1}$. They suggested that the high-velocity emission could arise from masers orbiting a massive central black hole, or ejected in a bipolar outflow. Using the Very Long Baseline Array (VLBA), \citet{Miy95} simultaneously observed the systemic and high-velocity water maser emission in NGC 4258, finding that the spatial distribution and line-of-sight velocities of the water masers trace a thin molecular ring in Keplerian rotation around a massive black hole of 3.6$\times10^7$~M$_{\odot}$ seen nearly edge-on.
The position-velocity (PV) diagram for the maser emission shows distinct Keplerian orbits (with deviations $<1\%$) defined by the high-velocity maser emission that arises on the ring diameter perpendicular to the line of sight, as well as a line traced by the systemic maser emission that arises from material on the inner edge of the amplifying ring, this linear dependence is a consequence of the change in the line-of-sight projection of the rotation velocity.

By monitoring both systematic and high-velocity water maser emission of NGC 4258 over periods of several years with different radio telescopes, a significant centripetal acceleration was observed only for the maser features near the galactic systemic velocity. 
The systemic maser features drift at a mean rate of $\sim$9~km~s$^{-1}$yr$^{-1}$ \citep{Has94,Gre95,Nak95,Bra00} while the high-velocity maser features drift by $\lesssim$1~km~s$^{-1}$yr$^{-1}$ \citep{Gre95}. In a recent spectroscopic study, \citet{Bra00} detected accelerations for the high-velocity features in the range of $-0.77$ to $0.38$~km~s$^{-1}$~yr$^{-1}$.
These measurements indicate that the systemic water masers lie within a relatively narrow range of radii, on the near side of the ring at the proximity of its inner edge, while the high-velocity water masers are located near the ring diameter \citep[between $-$13\grad6 and 9\grad3 of the mid-line,][]{Bra00}. In addition, the deviation of the high-velocity masers from a straight line passing through the systemic masers in the plane of the sky suggests that the rotating disk is slightly warped \citep{Her96,Her99,Her05}.

Previously, \citet{Wat94} demonstrated that the maser emission from a rapidly rotating, thin Keplerian ring viewed edge-on can reproduce the general features of the observed 22 GHz radiation from the nucleus of NGC 4258, including the high-velocity satellites. However, it is important to point out that their assumption of a uniform absorption coefficient within the amplifying ring results in a PV diagram for the most intense masers that is qualitatively different from the observed one. While their model predicts that the maser emission at velocities around the systemic velocity of the galaxy comes mainly from the outer edge of the ring, the observations indicate that this emission is actually coming from the inner parts of the truncated disk. 

In this paper we show that this discrepancy can be resolved if the absorption coefficient decreases with distance from the central mass. The model is presented in \S 2. The main results are described in \S 3. Finally, the conclusions are discussed in \S 4.

\section{Model}
We study the maser emission that arises from a thin, planar, gaseous ring in Keplerian rotation around a massive central object when it is observed edge-on.
The masing gas is located between $R_0$ and $R$, the inner and outer radii of the ring, respectively. 
For simplicity, we assume that the disk is transparent to the maser radiation at radii smaller than $R_0$ and greater than $R$, although the inner region is probably thermalized due to the higher gas density, and actually it would absorb a significant fraction of the maser radiation produced in the far side of the ring (see \S 4).
The absorption coefficient is assumed to follow a power law function of the distance from the central mass within the amplifying ring as, $\kappa=\kappa_0r^{-q}$. The distances are measured in units of $R$, and the velocities are measured in units of $v_{\mathrm{out}}$, the rotation velocity at the outer edge of the ring (see Figure 1).

For the case of an unsaturated maser and neglecting the spontaneous emission, the intensity of the maser radiation from a line of sight with impact parameter $y$ at a velocity $v_r$ is
\begin{equation}
I(v_r,y)=I_0e^{\tau(v_r,y)}\,,
\end{equation}
where the optical depth or gain along the line of sight is given by
\begin{equation}
\tau(v_r,y)=2\,\kappa_0\int^{x_{\mathrm{max}}}_{x_{\mathrm{min}}}(x^2+y^2)^{-q/2}\,\exp\Bigg[\frac{-(v-v_r)^2}{\Delta v_D^2}\Bigg]dx,
\end{equation}
with
\begin{displaymath}
x_{\mathrm{min}}=\cases{\sqrt{r_0^2-y^2} & \textrm{for} $0\leq\vert y\vert\leq r_0$\,,\cr
                        0 & \textrm{for} $r_0<\vert y\vert\leq 1$\,,\cr}
\end{displaymath}
and
\begin{displaymath}
x_{\mathrm{max}}=\sqrt{1-y^2}\,.
\end{displaymath}
The line-of-sight velocity component of the gas at the position $(x,~y)$ can be expressed as $v=y/(x^2+y^2)^{3/4}$. Here $I_0$ and $\Delta v_D$ are the background intensity and the Doppler width, respectively, which are supposed to be uniform inside the amplifying ring. The Doppler width $\Delta v_D$ is related with the FWHM of the velocity distribution of the emitting particles as $\Delta v_D=\mathrm{FWHM}/\sqrt{4\ln 2}$.

We have numerically solved equations (1) and (2), and we have also calculated the $y$-positions (impact parameters) of maximum maser intensity for each specific value of the velocity $v_r$. When we have found two local maxima, we have kept both. With this information, we have constructed the PV diagrams using the positions of the observer's line of sight with maximum emission at each velocity. This way to construct the PV diagrams was previously used by \citet{Usc05}.

We show the results using the following values for the model parameters which seem to be appropriate for modeling maser emission in the galaxy NGC 4258. The background intensity is  $I_0=1.3\times10^{-5}$~Jy~beam$^{-1}$, corresponding to a radio continuum source with a temperature of $10^6$~K \citep{Wat94}. The dimensionless inner radius $r_0=0.51$, using the estimated values for the inner and outer radii of 4.1 and 8.0~mas respectively, given by \citet{Miy95}.
The Doppler width $\Delta v_D=0.007v_{\mathrm{out}}$ which combined with an outer rotation velocity of 770~km~s$^{-1}$ \citep{Miy95}, gives a Doppler width $\simeq$~5~km~s$^{-1}$, similar to the value used by \citet{Wat94}. We have used some representative values of the exponent $q$, specifically $q=0,\;1/2,\;15/8,\;5$ for Models I, II, III and IV, respectively. In Model I, we study the simplest situation of a uniform absorption coefficient. In Model III, we choose $q=15/8$, that corresponds to the density dependence with the radius of an accretion disk, i.e., \citet{Fra92}. Finally, in Models II and IV, we explore two other different values of the exponent $q$ in order to study how it changes the results. In all the models, the value for the absorption coefficient $\kappa_0$ is mainly determined by the requirement that the intensity at the peak of the central component (13~Jy~beam$^{-1}$) is compatible with the observational data when the background intensity is 1.3$\times$10$^{-5}$~Jy~beam$^{-1}$. Other values of $I_0$, $r_0$, $\Delta v_D$, and $\kappa_0$ give qualitatively similar results.

We present the results in the next section; but let us first discuss briefly some important concepts in order to understand these results. In general, the observed emission at a given velocity coming from a specific position in a nebula has contributions of the whole material along the line of sight. However, when the velocity gradient along the line of sight is greater than the dispersion velocity (thermal or turbulent) of the emitting material, the main contribution to the emission is actually coming from a narrow region around the point with a line-of-sight velocity equal to the observation velocity. The estimated width of the region is $2l$, where $l$ is the \emph{correlation distance} defined as
\begin{equation}
l\equiv\frac{\Delta v_D}{\vert dv/dx\vert}\,,
\end{equation}
here $dv/dx$ is the line-of-sight velocity gradient. In this approximation, known as \textit{Sobolev's approximation} or the \textit{approximation of high velocity gradient}, the observed intensity is given by the following expression
\begin{equation}
I(v_r)=I_0(v_r)e^{-\tau(v_r)}+S(v_r)(1-e^{-\tau(v_r)})\,,
\end{equation}
where $I_0$ is the background intensity, $S$ is the source function and $\tau$ is the optical depth given by 
\begin{equation}
\tau(v_r)=\kappa(2l)\,,
\end{equation}
where $\kappa$ is the absorption coefficient.

For the case of maser emission, the value of $\kappa$ is intrinsically negative and $\tau$ is also negative, therefore the factor $e^{-\tau(v_r)}$ becomes an amplification factor. Because of this reason, the relative contribution at a given velocity of the correlation region is even more important with respect to the remainder of the emitting material than in the case of non-maser emission. Consequently, the approximation given by equations (4) and (5) is suitable for maser emission.

As shown in the next section, for a gaseous ring of inner radius $R_0$ and outer radius $R$ in Keplerian rotation and seen edge-on, the emission either comes preferentially from the inner or outer edges of the ring or from the line perpendicular to the line of sight and passing through the ring center. In the first two cases, it is easy to show that the expected PV diagram will be a straight line. When the emission comes from the outer edge, the slope of the straight line is equal to one (measuring the distances in units of the outer radius of the ring and the velocities in units of the rotation velocity at that point), whereas if the emission comes from the inner edge, the slope of the straight line is equal to $1/r_0^{3/2}$. On the other hand, when the emission arises from the line perpendicular to the line of sight, the PV diagram will be a curve with the form $1/y^{1/2}$, where $y$ is the impact parameter of the observation (see Figure 2).

\section{Results}
The PV diagrams for the maser emission peak are point-symmetric, consequently we only discuss positive velocities from now on (see Figures 3 and 4).
\begin{itemize}
\item Model I $(q=0)$ -- With this value of the exponent $q$, we are considering the simplest situation, a uniform or constant absorption coefficient. 
The strongest maser emission either comes mostly from the outer edge of the ring at velocities lower than 1, or from the mid-line of the ring perpendicular to the line of sight and passing through the central mass at greater velocities. The filled squares, circles, and triangles mark the regions of strongest maser emission at each velocity.
\item Model II $(q=1/2)$ -- The results are qualitatively similar to those of Model I.
\item Model III $(q=15/8)$ -- This value of exponent $q$ corresponds to the density dependence with radius of an accretion disk ($\rho\propto r^{-15/8}$). The strongest maser emission either comes mainly from the inner edge of the ring at low velocities (velocities near the systemic velocity), or from the outer edge at velocities close to 1. On the other hand, at velocities greater than 1, the most intense emission comes predominantly from the mid-line of the ring perpendicular to the line of sight.
\item Model IV $(q=5)$ -- The strongest maser emission comes mainly from the inner edge of the ring at velocities lower than 1. At greater velocities, the most intense emission can either come mainly from the inner edge or from the mid-line of the ring perpendicular to the line of sight.
\end{itemize}

In summary, from the results of Models I--IV (see Figure 3), we found that the most intense maser emission can be around the inner or outer edges of the ring, or the mid-line of the ring perpendicular to the line of sight depending on the velocity and also on the value of $q$. In fact, the PV diagrams are qualitatively different when $q<1$ or $q>1$. In the first case, for $q<1$ (including the simplest situation with a uniform absorption coefficient, $q=0$) and $v_r<1$, the PV diagram corresponds to a straight line with slope 1; for $v_r>1$, the diagram corresponds to a Keplerian curve. In the second case, for $q>1$ and $v_r<1$, the PV diagram corresponds to a straight line with a slope that depends on the inner radius of the ring. At velocities close to 1 the slope changes to 1; for $v_r>1$, the diagram corresponds to a Keplerian curve and also a straight line with a slope that depends on the inner radius under circumstances such as in Model IV.

It is also important to realize that when $q>1$, the optical depth or gain presents two local maxima within a certain velocity range. Either local maxima may be a global maximum. For $v_r<1$, the local maximum can be either at the inner and/or outer edges of the ring, while for $v_r>1$, they are located at the inner edge and/or mid-line of the ring (see Figure 5).

As shown in the top panels of Figure 5 (Model III, $q=15/8$), the relative difference between the two local maxima is not very significant. However, when the value of $q$ is higher (like in Model IV, $q=5$ shown in the bottom panels) the relative difference becomes more important.

In order to estimate the velocity $v_c$ at which the global maximum of the optical depth changes its locus, we have calculated analytical approximations for the largest value of the optical depth or gain that corresponds to the maximum intensity at the inner and outer edges of the ring, and also at the mid-line of the ring perpendicular to the line of sight. 
The detailed calculations are presented in the Appendix.
The following equations give the local maximum depth as a function of the velocity in each neighborhood
\begin{eqnarray}
\tau(v_r)\simeq\frac{4}{3}\sqrt{\pi}\kappa_0\Delta v_D\frac{r_0^{1-q}}{v_r\sqrt{1-r_0v_r^2}}\quad \textrm{inner edge}\,,\\
\tau(v_r)\simeq\frac{4}{3}\sqrt{\pi}\kappa_0\Delta v_D\frac{1}{v_r\sqrt{1-v_r^2}}\quad \textrm{outer edge}\,,\\
\tau(v_r)\simeq\frac{4}{\sqrt{3}}\kappa_0\sqrt{\Delta v_Dv_r^{4q-5}}\quad \textrm{mid-line}\,.
\end{eqnarray}

The velocity $v_c$ is estimated by combining equations (6) and (7), or equations (6) and (8) according to the value of $v_c$ (when $v_c<1$ or $v_c>1$, respectively). The results are
\begin{mathletters}
\begin{eqnarray}
v_c={\Bigg[\frac{r_0^{2(1-q)}-1}{r_0^{2(1-q)}-r_0}\Bigg]}^{1/2} \; \textrm{for}\; v_c<1,\\
\frac{\pi}{3}\Delta v_Dr_0^{2(1-q)}-v_c^{4q-3}+r_0v_c^{4q-1}=0\: \textrm{for}\; v_c>1,
\end{eqnarray}
\end{mathletters}
which are presented in Figure 6 for some representative values of the exponent $q$.

The bottom plot of Figure 6 shows $v_c$ as function of the inner radius $r_0$ for different values of $q$, from equation (9a). For $q<1$, there is no solution to equation (9a). When $q=1$, $v_c=0$ for any value of $r_0$. That is, the optical depth has a maximum and its locus is around the outer edge of the ring, and $v_c$ is meaningless as we have defined it.
When $q>1$, the optical depth presents two local maxima, and $v_c$ is different from zero and its value depends on $r_0$. This velocity corresponds to the value at which the locus of the global maximum changes from the inner edge to the outer edge of the ring. As a consequence, there is a slope change in the PV diagrams at velocities lower than the rotation velocity at the outer edge of the ring. For instance, when $q=15/8$ and $r_0=0.51$, the slope change occurs at $v_c=0.906$. In other words, the locus of the global maximum of the optical depth changes from the inner to the outer edge of the ring at this velocity $v_c$.

The top plot of Figure 6 also shows $v_c$ as function of the inner radius $r_0$ using specific values of $q$ and $\Delta v_D$ in equation (9b); in this case, 5 and 0.007$v_{\mathrm{out}}$, respectively. As an example, when $\Delta v_D=0.007v_{\mathrm{out}}$, $q=5$ and $r_0=0.51$, then $v_c=1.077$. Stated differently, at that velocity $v_c$, the largest value of both the optical depth and the intensity changes its locus from the inner edge to the mid-line of the ring perpendicular to the line of sight.

The remarkable water maser emission in the nucleus of the galaxy NGC 4258 traces a PV diagram where the detected emission around the systemic velocity of the galaxy comes from the inner edge of the amplifying ring; this emission delineates a straight line just as the straight line that connects points C and D in Figure 2 \citep[see Figure 3 of][]{Miy95}. According to our model results, this implies that the absorption coefficient within the molecular ring of NGC 4258 is not uniform, instead it must be a decreasing function of the distance from the central mass, i.e., $\kappa=\kappa_0r^{-q}$ with $q>1$. Moreover, the observed red/blue-shifted emission at high velocities that arises from the mid-line of the ring perpendicular to the line of sight traces a Keplerian curve such as is indicated by the model results (see Figure 7). Simply stated, when $q>1$ the PV diagram is qualitatively similar to the one observed for the water maser emission detected in the nucleus of NGC 4258. 

As an example, in Figure 7 we show a comparison between the results of Model III ($q=15/8$) and the water maser emission in NGC 4258. The detected emission arises from the inner edge of the amplifying ring and the mid-line perpendicular to the line of sight. The locus of the observed maser emission coincides with the locus of the most intense maser emission as indicated by the sizes of the circles in Figure 7. The model results indicate that there is emission coming from the outer edge of the ring at velocities close to 1, nevertheless the sizes of the circles indicate that this emission is very weak. Maybe maser emission is not detected from this locus for this reason.

Additionally, our model results also indicate that the intense maser emission at the inner edge of the ring extends neither to velocities very different from the systemic velocity nor to impact parameters very different from zero, as is indicated by the size of the circles in the PV diagram shown in Figure 7. Furthermore, according to the size of the circles, the other locus of intense maser emission is the mid-line of the ring perpendicular to the line of sight, precisely the locus of the red/blue-shifted maser emission at high velocities that describes Keplerian curves in the PV diagram.

\section{Discussion and Conclusions}
In our model, we have assumed that the gas in the region inside the masing ring is transparent to the maser radiation. This implies that the most intense maser emission at low velocities (velocities near the systemic velocity) comes mainly from the outer edge of ring (for $q<1$) or from the inner edge (for $q>1$), either the near or far side of the ring, as is indicated in Figure 4. Measurements of positive acceleration of the maser emission around the systemic velocity show that this emission certainly comes from the near side of the ring at the proximity of its inner edge \citep[e.g.,][]{Gre95}. If we suppose that the gas inside the masing ring is thermalized probably due to its higher density then an important fraction of the maser emission from the backside of the ring would be absorbed and the detected emission would come from the front side of the ring at the outer or inner edge depending on the value of $q$. For instance, considering absorption and emission from the gas located inside the masing ring, the difference in the intensity for a line of sight that passes through both the inner absorbing region and the front side of the masing ring from the intensity for a line of sight that passes through both the backside of the masing ring and the inner absorbing region is $S(1-e^{-\tau_2})(e^{\tau_1}-1)$ where $S$ is the source function of the gas inside the masing ring, $\tau_2$ is the optical depth in this region, and $\tau_1$ is the optical depth for the front side of the ring. If $\tau_2>>\tau_1$, then the detected emission would be the radiation amplified by the front side of the masing ring.

Also, we have made a simplifying assumption about the geometry of the masing ring in NGC 4258, considering that the amplifying ring is strictly flat. Despite the observations indicate an apparent warp in the maser distribution of this galaxy, \citet{Kar99} presented a model in which the disk does not require to be physically warped in order to the masing gas become exposed to the central continuum radiation. In this scenario, dusty clouds provide the shielding of the high-energy continuum, which is required for the gas to remain molecular. They found that a flat-disk model of the irradiated ring could be applied to a source like NGC 4258 only if the water abundance is higher than the value implied by equilibrium photoionization-driven chemistry. A very important result from their study (based on radiative and kinematic considerations) was that, even if the disk in NGC 4258 is warped, the maser-emitting gas must be clumpy, instead of homogeneous as in the scenario previously proposed by \citet{Neu95}. 

An important result of our model shows that the assumption, commonly used, that considers a uniform or constant absorption coefficient within the masing ring in Keplerian rotation around the nucleus of NGC 4258 is not appropriate. For example, \citet{Wal98} supposed that $\kappa$ was constant considering that the locus of the maser emission from NGC 4258 was determined mainly by the velocity gradients in a Keplerian velocity field indicating some uniformity of $\kappa$, at least on length scales comparable to the coherence or correlation length resulting from the Keplerian velocity gradients. On the contrary, from our analysis, we conclude that a constant absorption coefficient would result in a PV diagram qualitatively different from the observed one, since the most intense maser emission would come predominantly from a narrow region close to the outer edge of the ring instead of a narrow region close to the inner edge of the ring, as indicated by the observations. Necessarily, the absorption coefficient must be a decreasing function of distance from the central mass (i.e., $\kappa=\kappa_0r^{-q}$ with $q>1$) to have significant emission coming from the inner edge of the amplifying ring and hence explain the form of the PV diagram delineated by the water masers in NGC 4258.  

When comparing our edge-on disk model with the observations of NGC 4258, it is clear that we need a $\kappa\propto r^{-q}$ radial dependence for the absorption coefficient with $q>1$ (so as to favour the emission from the inner edge of the disk, see above) in order to reproduce the observations. In reality, the fact that the disk of NGC 4258 is warped introduces geometrical effects which might favour the inner disk edge emission (over the one of
the outer edge). One will need to compute more complex, 3D transfer models to see whether or not these geometrical effects are sufficient to explain the PV diagrams of the NGC 4258 masers without introducing the radially dependent absorption coefficient which is required by the edge-on disk models described in the present paper.

\acknowledgments

J. C. and A. C. R. acknowledge support from CONACyT grants 41320 and 43103, and DGAPA-UNAM. L. U. acknowledges support from DGAPA-UNAM. We sincerely thank J. M. Torrelles and Y. G\'omez for useful comments, which contributed to improve an earlier version of this manuscript. L. U. gives special thanks to M. R. Pestalozzi and M. Elitzur for valuable comments on this work.
We also thank an anonymous referee for helpful comments on the manuscript.

\appendix
\section{Analytical approximations for the optical depth}
In this appendix, we describe how to obtain the analytical approximations for the optical depth given by equations (6)--(8).

First, we define $w=(v-v_r)/\Delta v_D$ then we can change the variables in equation (2) and rewrite it as
\begin{equation}
\tau(v_r,y)=2\kappa_0\int_{w_{\mathrm{min}}}^{w_{\mathrm{max}}}(x^2+y^2)^{-q/2}\frac{dx}{dw}\exp(-w^2)dw\,,
\end{equation}
where
\begin{displaymath}
w_{\mathrm{min}}=\frac{y/r_0^{3/2}-v_r}{\Delta v_D}\,,
\end{displaymath}
and
\begin{displaymath}
w_{\mathrm{max}}=\frac{y-v_r}{\Delta v_D}\,.
\end{displaymath}
Note that $w_{\mathrm{min}}>w_{\mathrm{max}}$ since $r_0\leq 1$.
Using the expressions for the line-of-sight velocity $v=y/(x^2+y^2)^{3/4}$ and the previously defined variable $w=(v-v_r)/\Delta v_D$, we can write $x=\Big[{{(y/(v_r+w\Delta v_D))}^{4/3}-y^2}\Big]^{1/2}$.
At zero order around $w=0$ we obtain
\begin{equation}
(x^2+y^2)^{-q/2}\frac{dx}{dw}\simeq-\frac{2\Delta v_D(y/v_r)^{2(2-q)/3}}{3v_r\sqrt{(y/v_r)^{4/3}-y^2}}\,.
\end{equation}
Additionally
\begin{equation}
\int^{w_{\mathrm{max}}}_{w_{\mathrm{min}}}\exp(-w^2)dw=\frac{\sqrt{\pi}}{2}\Bigg[\mathrm{erf}(w_{\mathrm{max}})-\mathrm{erf}(w_{\mathrm{min}})\Bigg]\,,
\end{equation}
where $\mathrm{erf}(w)$ is the error function, defined as $\mathrm{erf}(w)\equiv (2/\sqrt{\pi})\int_{0}^{w}\exp(-t^2)dt$.
Finally, substituting equations (A2) and (A3) into (A1), we obtain the approximation for the optical depth
\begin{equation}
\tau(v_r,y)\simeq\frac{2\sqrt{\pi}\kappa_0\Delta v_D(y/v_r)^{2(2-q)/3}}{3v_r\sqrt{(y/v_r)^{4/3}-y^2}}\Bigg[\mathrm{erf}(w_{\mathrm{min}})-\mathrm{erf}(w_{\mathrm{max}})\Bigg]\,.
\end{equation}

Around the inner edge of the ring, $v_r\simeq y/r_0^{3/2}$ and the maximum value of $[\mathrm{erf}(w_{\mathrm{min}})-\mathrm{erf}(w_{\mathrm{max}})]=2$. Substituting these approximations into equation (A4), we obtain equation (6). Similarly, around the outer edge of the ring, $v_r\simeq y$, and the maximum value of $[\mathrm{erf}(w_{\mathrm{min}})-\mathrm{erf}(w_{\mathrm{max}})]$ also equals 2. Then, equation (A4) reduces to equation (7) for the optical depth at the outer edge of the ring.

In order to find an approximation for the local maximum of optical depth at the mid-line of the ring perpendicular to the line of sight, we expand the expression for the velocity along the line of sight around $x=0$ to obtain
\begin{equation}
v\simeq\frac{1}{y^{1/2}}-\frac{3}{4}\frac{x^2}{y^{5/2}}\,,
\end{equation}
therefore
\begin{equation}
v-\frac{1}{y^{1/2}}=-\frac{3}{4}\frac{x^2}{y^{5/2}}=-\Delta v_D\,,
\end{equation}
and thus
\begin{equation}
x=\frac{2}{3^{1/2}}\Delta v_D^{1/2}y^{5/4}\,.
\end{equation}
Then, using \textit{Sobolev's approximation} 
\begin{equation}
\tau=\kappa_0(x^2+y^2)^{-q/2}(2x)\,,
\end{equation}
and substituting equation (A7) into (A8), we obtain the following expression
\begin{equation}
\tau\simeq\kappa_0\Big(\frac{4}{3}\Delta v_Dy^{5/2}+y^2\Big)^{-q/2}\frac{4}{3^{1/2}}\Delta v_D^{1/2}y^{5/4}\,,
\end{equation}
since $y\ll 1$ and $\Delta v_D$ is small, then $\frac{4}{3}\Delta v_Dy^{5/2}\ll y^2$, hence
\begin{equation}
\tau\approx\frac{4}{3^{1/2}}\kappa_0\Delta v_D^{1/2}y^{5/4-q}\,,
\end{equation}
considering that $y=v_r^{-2}$, we finally obtain the approximation for the local maximum of the optical depth at the mid-line of the ring given by equation (8).

{}

\clearpage

\begin{figure}
\epsscale{0.8}
\plotone{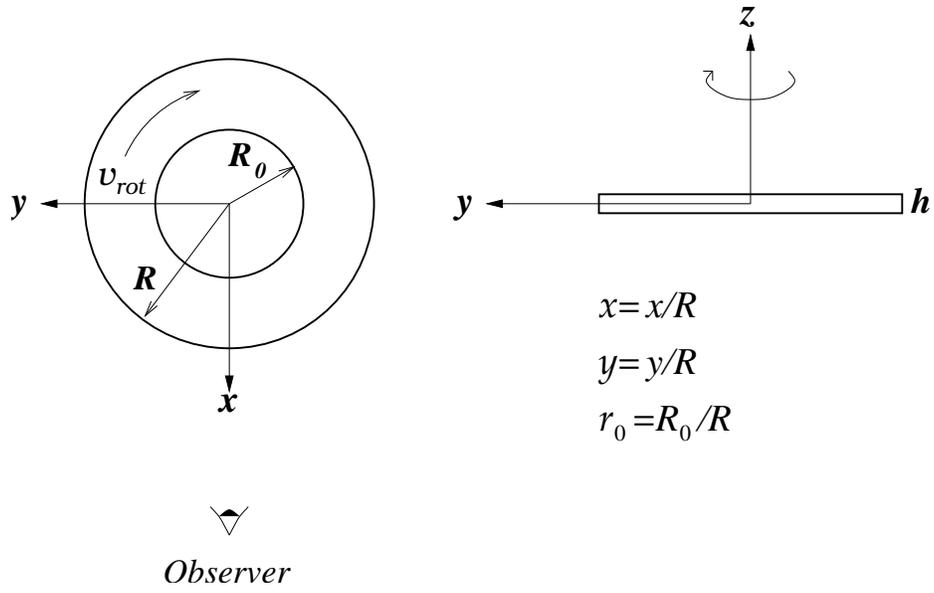}
\caption{Schematic diagram of a gaseous disk in Keplerian rotation. The masing gas exists between radii $R_0$ and $R$. At radii smaller than $R_0$ and greater than $R$ the disk is transparent to the maser radiation. The observer is on the plane of the disk. All the distances are measured in units of $R$, the outer radius of the amplifying ring; therefore the variables $x$, $y$, and $r_0$ are dimensionless. \label{fig1}}
\end{figure}

\clearpage

\begin{figure}
\epsscale{0.6}
\plotone{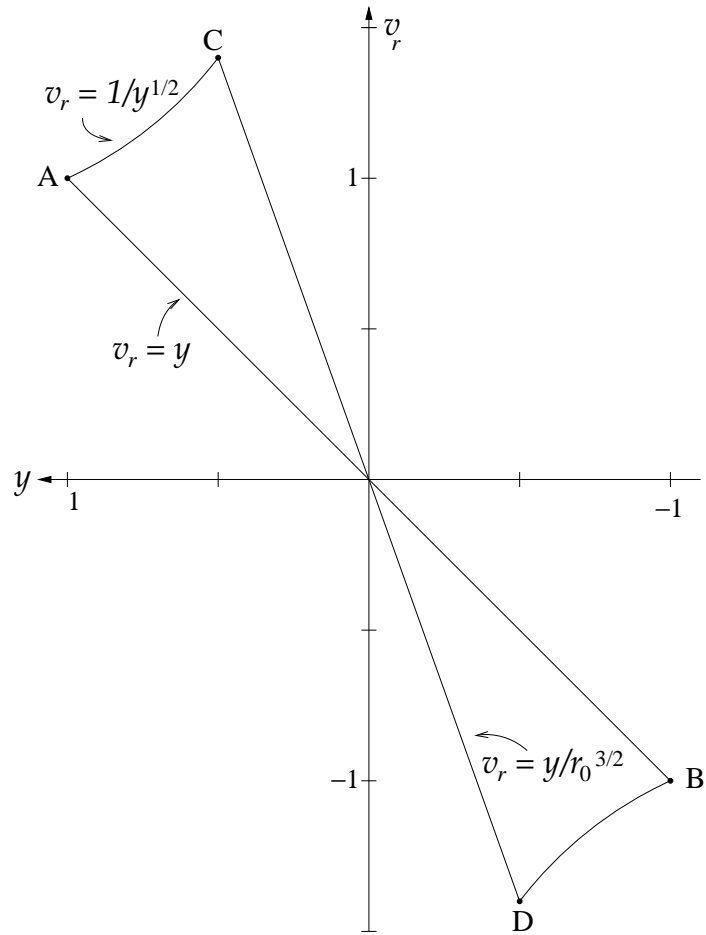}
\caption{PV diagram for the maser emission of a gaseous ring with inner radius $R_0$ and outer radius $R$ in Keplerian rotation observed edge-on. The straight line that connects points A and  B has a slope equal to 1, while the straight line that connects points C and D has a slope equal to $1/r_0^{3/2}$. \label{fig2}}
\end{figure}

\clearpage
\begin{figure}
\epsscale{0.8}
\plotone{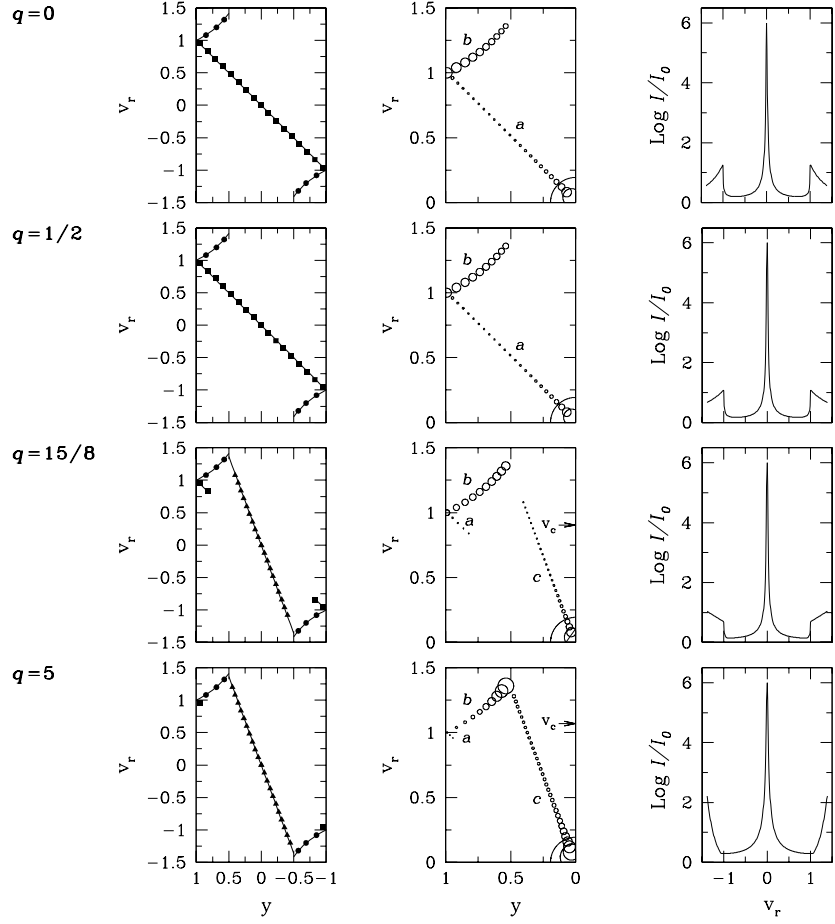}
\caption{From top to bottom results of Models I, II, III, and IV. The left panels show PV diagrams for the maser emission peak. The filled squares, circles, and triangles represent the strongest maser emission which is coming from regions \textit{a}, \textit{b}, and \textit{c}, respectively, indicated in Figure 4. The straight lines or curves represent the velocity dependences of the regions where this emission arises. The central panels show PV diagrams for the maser emission peak. Because of the point-symmetric shape of these diagrams, only positive velocities are shown. The radii of the open circles are proportional to the maximum maser intensity at each position and velocity. The right panels show the logarithm of the ratio between the maximum intensity and the background intensity as a function of the velocity. \label{fig3}}
\end{figure}

\clearpage

\begin{figure}
\epsscale{1.0}
\plotone{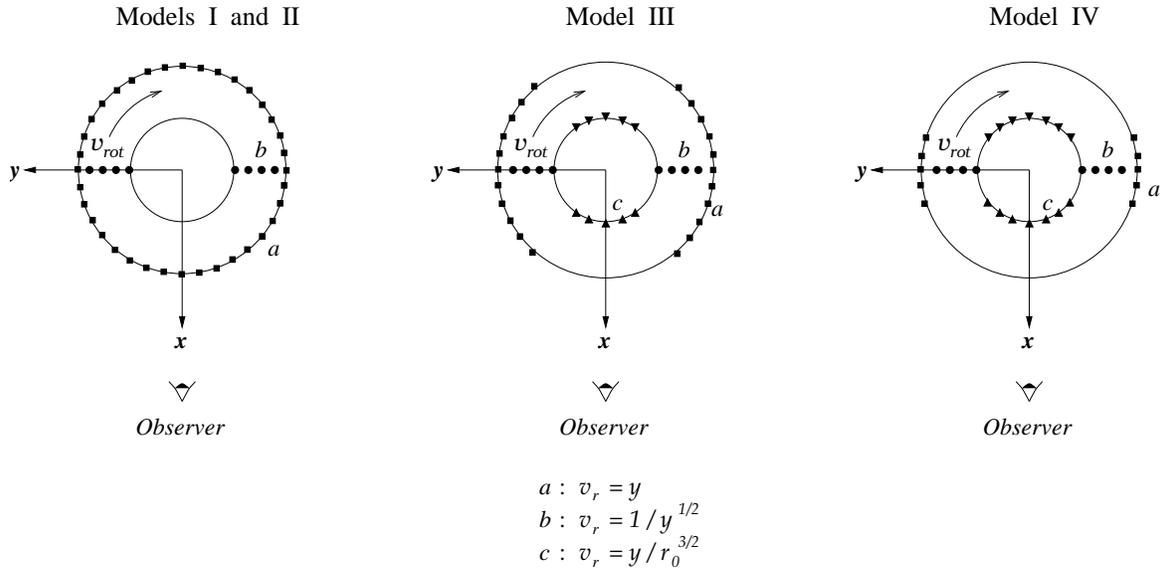}
\caption{Schematic representation of the results of Models I, II, III, and IV. The filled squares, circles, and triangles represent the most intense maser emission that is coming from regions \textit{a}, \textit{b}, and \textit{c}, respectively. These regions are very narrow because the correlation distance is very small; that is, the width $\Delta v_D$ is much smaller than the line-of-sight velocity gradient. Besides, the exponential amplification of the intensity emphasizes small changes in the optical depth.\label{fig4}}
\end{figure}

\clearpage

\begin{figure}
\epsscale{0.75}
\plotone{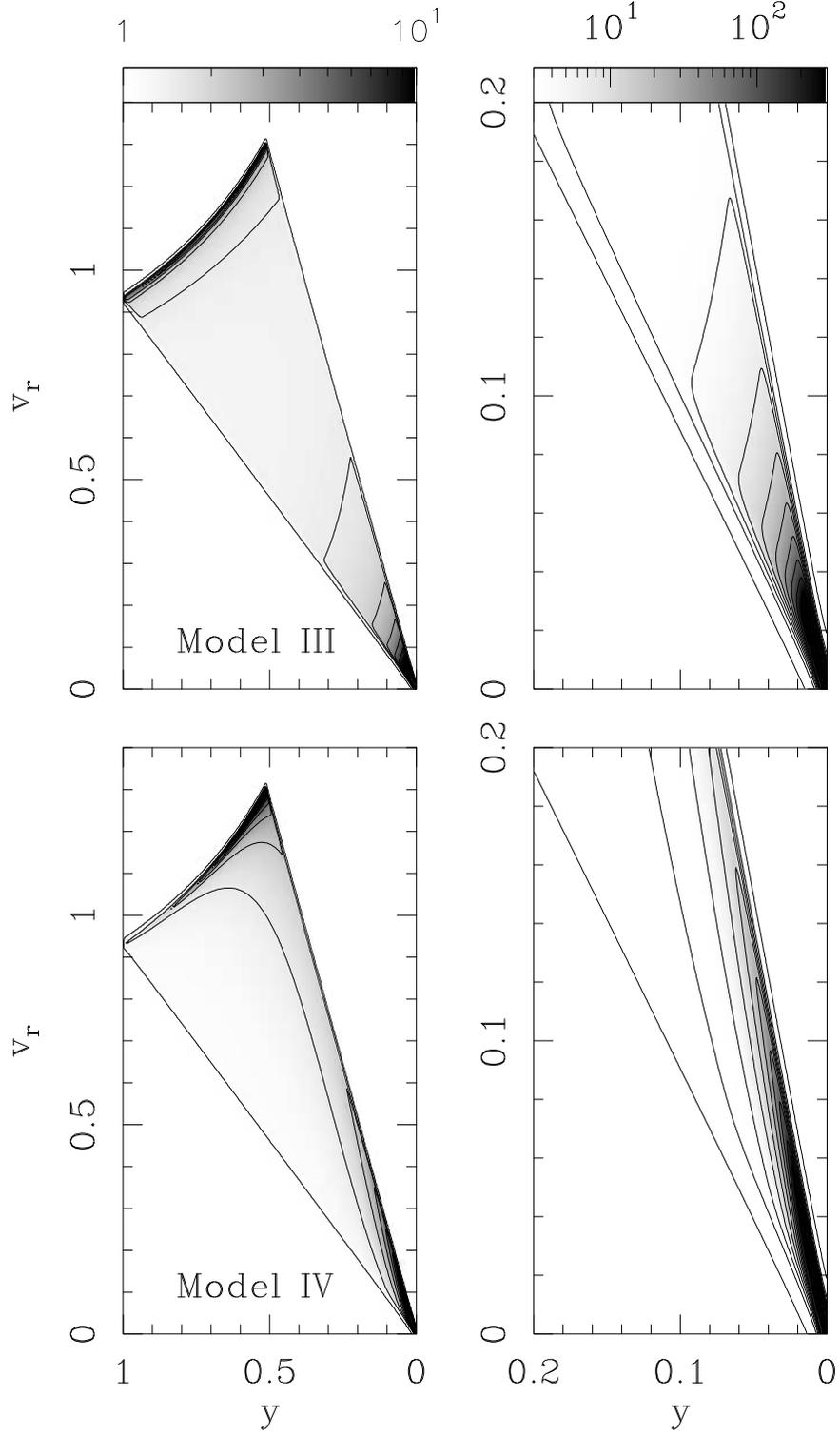}
\caption{\textit{Left}: PV diagrams for the maser emission in grey scale with intensity contours overlaid. The darker regions show the locus of the strongest emission in these diagrams, which potentially could be detected, depending on the sensitivity cutoff of the observations. \textit{Right}: Close-up to the PV diagrams showing the maser emission at low velocities. \label{fig5}}
\end{figure}
\clearpage

\begin{figure}
\epsscale{0.8}
\plotone{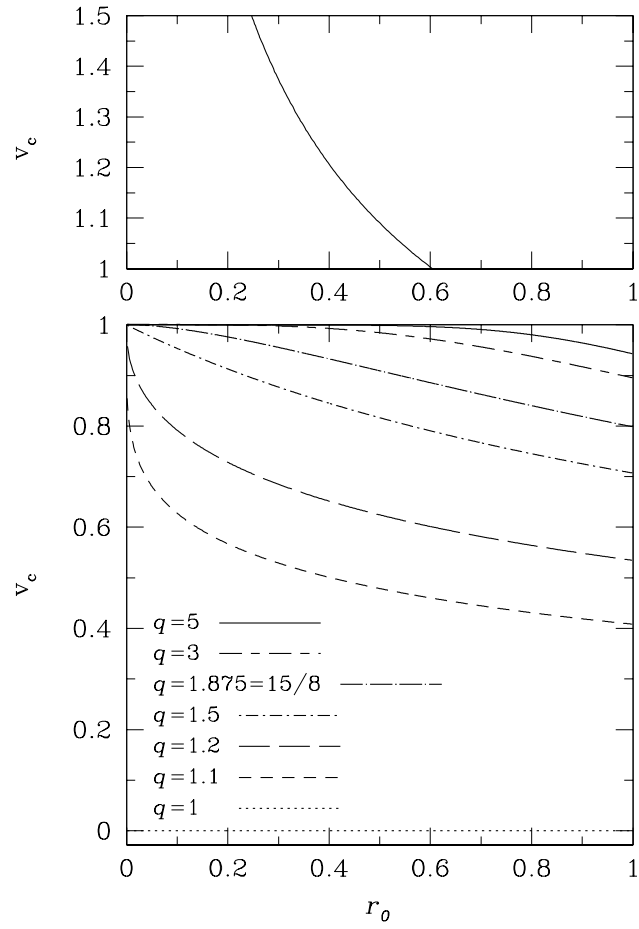}
\caption{$v_c$ as function of the inner radius of the ring, $r_0$. \textit{Bottom}: For $v_c<1$, it is computed from equation (9a) for some representative values of $q$. For $q<1$, there is no solution to equation (9a). \textit{Top}: For $v_c>1$, it is computed from equation (9b). This is the solution to equation (9b), using $\Delta v_D=0.007v_{\mathrm{out}}$ and $q=5$. \label{fig6}}
\end{figure}

\clearpage

\begin{figure}
\epsscale{0.8}
\plotone{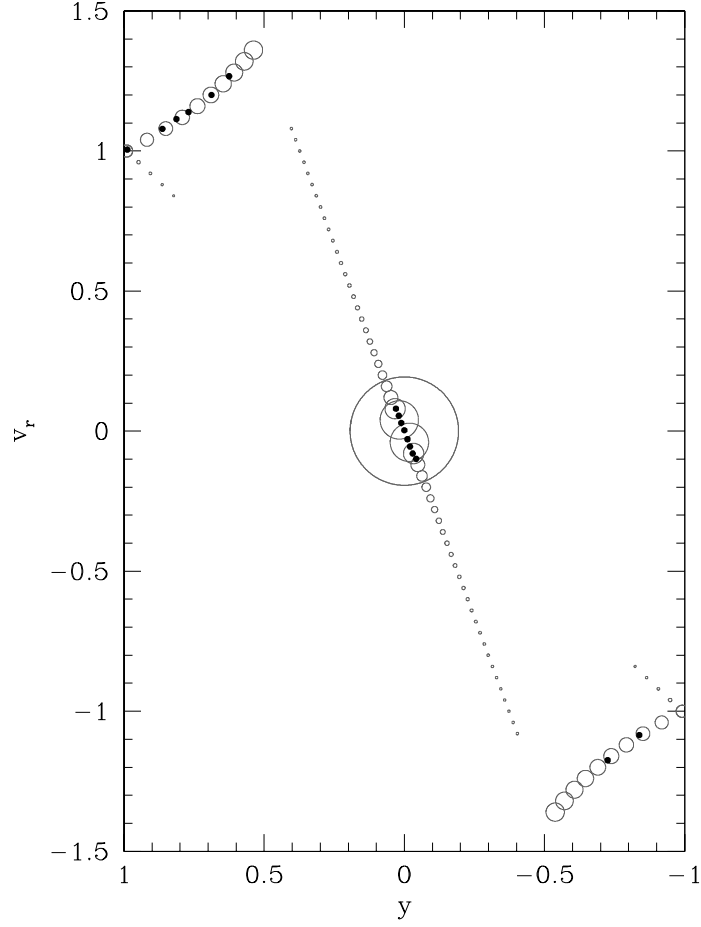}
\caption{Comparison between the calculated PV diagram for the maser emission peak in Model III ($q=15/8$) and the PV diagram delineated by the water masers observed in NGC 4258 (Miyoshi et al. 1995). The radii of the open circles are proportional to the maser intensity at each position and velocity. The dots represent the observed maser spots in NGC 4258. We have subtracted the ring systemic velocity of 476 km s$^{-1}$ from the observed local standard of rest velocity of the maser spots in order to compare the observed PV diagram with the modeled one. The positions and velocities are in units of the outer radius of the ring (8 mas) and the rotation velocity at the outer edge of the ring (770 km s$^{-1}$), respectively. \label{fig7}}
\end{figure}

\end{document}